\documentclass[prb,twocolumn,showpacs,amsmath,amssymb,superscriptaddress,floatfix]{revtex4-2}

\usepackage{graphicx}
\graphicspath{{}}
\usepackage{dcolumn}
\usepackage{bm}
\usepackage{amssymb,mathtools}
\usepackage{booktabs}
\usepackage{siunitx}
\usepackage{color}
\usepackage{microtype}
\usepackage[T1]{fontenc}
\usepackage{lmodern} 
\usepackage{times}
\usepackage{enumitem}

\usepackage{physics}
\usepackage{upgreek}

\allowdisplaybreaks
\usepackage{placeins}

\usepackage[plainpages=falsepdfpagemode=UseNone,pdfstartview=FitBH]{hyperref}
\hypersetup{
colorlinks = true,
linkcolor = [rgb]{0.87,0.18,0.22},
citecolor = [rgb]{0.0,0.60,0.32},
urlcolor = [rgb]{0.20, 0.30, 0.54}}

\usepackage{sidecap}

\renewcommand{\tablename}{Table}
\makeatletter\renewcommand{\fnum@table}[1]{\tablename~\thetable.}\makeatother

\makeatletter

\begin{document}
\title{Quantum criticality and emergent orders in the spin-1 bilinear-biquadratic-Kitaev chain}

\author{Zhiling Wei}
\affiliation{National Laboratory of Solid State Microstructures and Department of Physics, Nanjing University, Nanjing 210093, China}

\author{Zhengzhong Du}
\affiliation{National Laboratory of Solid State Microstructures and Department of Physics, Nanjing University, Nanjing 210093, China}

\author{Xiaodong Cao}
\affiliation{Suzhou Institute for Advanced Research, University of Science and Technology of China, Suzhou 215123, China}
\affiliation{School of Artificial Intelligence and Data Science, University of Science and Technology of China, Hefei 230026, China}

\author{Wen-Long You}
\email[]{wlyou@nuaa.edu.cn}
\affiliation{College of Physics, Nanjing University of Aeronautics and Astronautics, Nanjing 211106, China}
\affiliation{Key Laboratory of Aerospace Information Materials and Physics (NUAA), MIIT, Nanjing 211106, China}

\author{Yi Lu}
\email[]{yilu@nju.edu.cn}
\affiliation{National Laboratory of Solid State Microstructures and Department of Physics, Nanjing University, Nanjing 210093, China}
\affiliation{Collaborative Innovation Center of Advanced Microstructures, Nanjing University, Nanjing 210093, China}

\date{\today}
\begin{abstract}
Higher-spin quantum magnets with competing interactions offer a rich platform for exploring quantum phases that transcend the paradigms of spin-1/2 systems, owing to their enlarged local Hilbert spaces and the emergence of multipolar correlations. We investigate a one-dimensional spin-1 chain where quadrupolar order is promoted by two distinct mechanisms: conventional bilinear-biquadratic exchange and bond-directional antiferromagnetic Kitaev frustration. Using density matrix renormalization group calculations, we determine the complete ground-state phase diagram and uncover two emergent phases induced by the Kitaev interaction: a Kitaev nematic phase and a Kitaev-dimer phase. The Kitaev nematic phase emerges from a fragile biquadratic dimer state via a continuous quantum phase transition in the Ising universality class. The Kitaev dimer phase spontaneously breaks a screw symmetry to favor either $x$- or $y$-spin bonding, forming a gapped state that coexists with a crystalline order of alternating $\mathbb{Z}_2$ fluxes.
\end{abstract}
\maketitle

\section{\label{sec:intro} Introduction}

Higher-spin ($S \ge 1$) quantum magnets have long been recognized as a fertile arena for emergent phenomena that are absent in their spin-1/2 counterparts. A paradigmatic example is the symmetry-protected topological (SPT) Haldane phase of the spin-1 Heisenberg chain, characterized by non-local string order and fractionalized edge states, features that stand in sharp contrast to the gapless, critical behavior of the spin-1/2 chain and have been foundational in the modern understanding of SPT phases~\cite{1983Haldane,1988Haldane,1992KTtransformation,Kennedy1992Z2Z2,Pollmann2010ESofSPT}.
An equally striking contrast arises in the context of Kitaev models with bond-dependent exchange frustration. For spin-1/2 degrees of freedom, the model is exactly solvable on the honeycomb lattice, where its ground state is a $\mathbb{Z}_2$ quantum spin liquid (QSL) composed of itinerant Majorana fermions coupled to static fluxes, representing a landmark example of fractionalization~\cite{Kitaev2006,Jackeli2009,Chaloupka2010,Plumb2014,Trebst2022review}. When extended to higher spins, however, the Kitaev model becomes non-integrable, and its ground-state properties deviate markedly. While flux conservation persists, the nature of fractionalization changes. The gauge charges become bosonic for integer spins~\cite{Read1991}, allowing for flux condensation and destabilizing the spin-liquid phase~\cite{Senthil2004DQCP}.
Consequently, recent theoretical and numerical studies support the picture that symmetry-breaking quadrupolar order, known as spin-nematicity, rather than spin-liquid behavior, emerges as the dominant motif in spin-1 Kitaev systems~\cite{2008SpinShoneycombKitaev,Xu2020,2020PhysRevResearch.2.022047,2020PhysRevResearch.2.023361,2021PhysRevResearch.3.013216,WLYou2020HK,2022WLYouManybodyscar,2023HKSIA}.

This positions the spin-1 system at an intriguing juncture. On one hand, the well-known bilinear-biquadratic (BBQ) Hamiltonian provides a canonical platform for realizing spin nematic quadrupole-ordered phases through conventional exchange interactions~\cite{2006BBQchain,2015QMCBBQ,2018iPEPSBBQ,2019nematicBBQ,2020nematicBBQ}. On the other hand, the spin-1 Kitaev interaction provides a distinct route to nematicity via bond-directional frustration. This overlap of nematic tendencies raises a natural question: What is the outcome of the interplay between these two mechanisms for nematicity, and how do their respective phases evolve and compete under mutual influence?

\begin{figure}[t]
  \includegraphics[width=\linewidth]{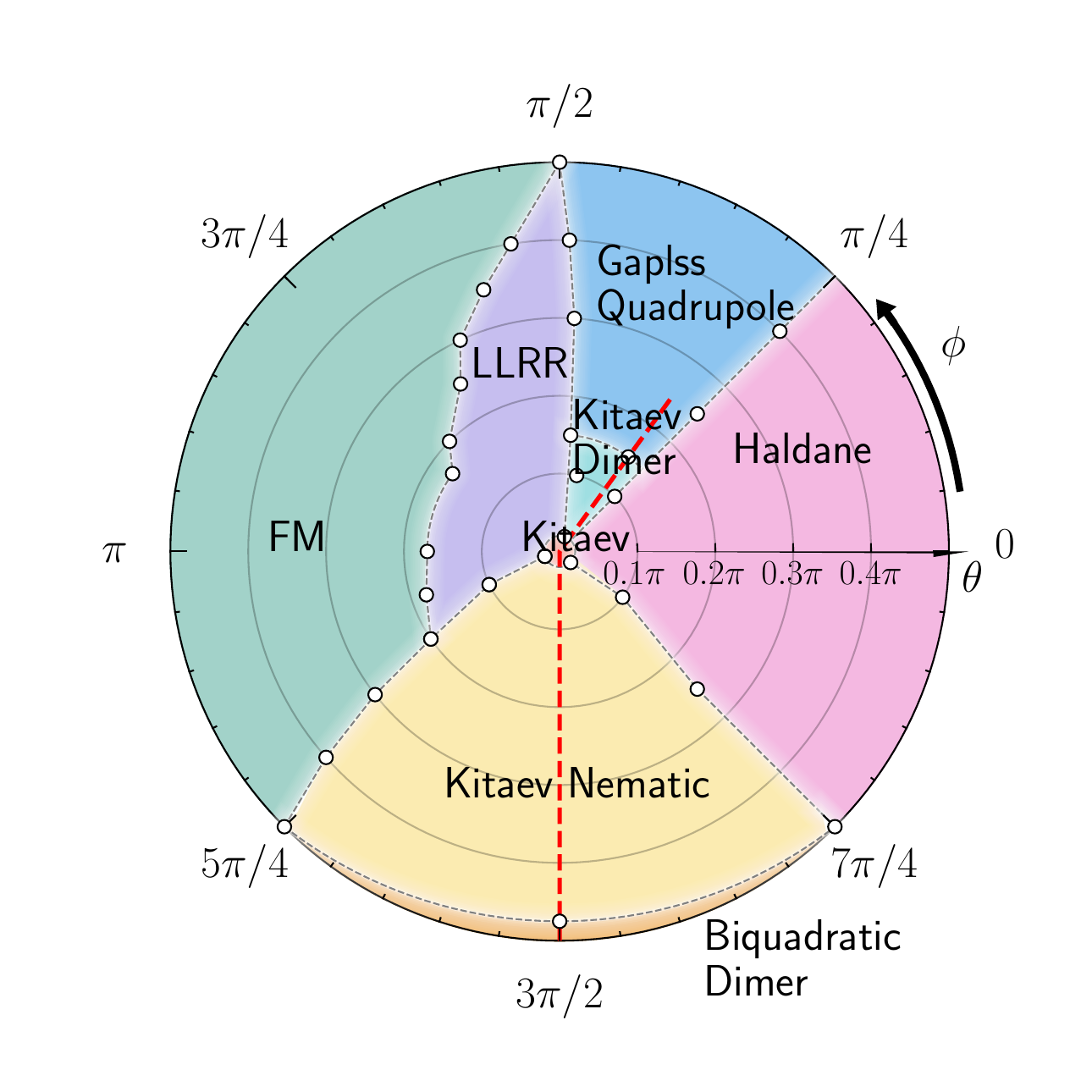}
  \caption{\label{Fig_PD}
  Phase diagram of the BBQK model~\eqref{eq:HBBQK} with antiferromagnetic Kitaev interaction $K\geq0$, parameterized as in Eq.~(\ref{eq:para}). The north pole ($\theta=0$) corresponds to the pure Kitaev model, while the equator ($\theta=\pi/2$) represents the BBQ model. Phase boundaries are determined by interpolating parameter-space points where DMRG calculations were performed (white dots).
  }
\end{figure}

In this paper, we address this question by numerically determining the ground-state phase diagram of the one-dimensional (1D) spin-1 BBQ-Kitaev (BBQK) chain [Eq.~\eqref{eq:HBBQK}] using the density matrix renormalization group (DMRG) method~\cite{1992White,2005reviewDMRG,SCHOLLWOCK201196}. Our main findings, summarized in Fig.~\ref{Fig_PD}, reveal two novel phases induced by the Kitaev interaction in addition to those known from the individual BBQ~\cite{2006BBQchain,1993TXiangBBQchain,1995BBQchain,1997Itoi,2011BBQwithZeemanOptical,2011BBQwithZeeman,2017BBQwithZeeman} and Kitaev~\cite{WLYou2020HK,2022WLYouManybodyscar,2023HKSIA} models: a \emph{Kitaev nematic phase} and a \emph{Kitaev dimer phase}. The former emerges from the fragile biquadratic dimer phase via a quantum critical point in the Ising universality class, while the Kitaev dimer phase spontaneously breaks a combined screw symmetry and simultaneously realizes a $\mathbb{Z}_2$ flux crystal with an alternating pattern. Our work thus provides a comprehensive characterization of the competition between exchange-driven and frustration-induced quadrupolar order in spin-1 systems.

The remainder of this paper is organized as follows. Section~\ref{sec:model} introduces the BBQK model, reviews its known properties in certain parameter limits, and discusses the invariants of the Kitaev interaction and potential symmetry-breaking scenarios. Section~\ref{sec:methods} details the numerical methods and diagnostic tools used in our study, including quantum information measures such as the entanglement spectrum and fidelity susceptibility~\cite{WLYou2007FideSusc}. In Sec.~\ref{sec:result}, we present our primary numerical results, identifying and characterizing two emergent nontrivial phases, the Kitaev dimer and Kitaev nematic phases, that arise from the interplay between the Kitaev and biquadratic interactions. Finally, we summarize our findings and conclude in Sec.~\ref{sec:conclusion}.

\section{\label{sec:model} 1D BBQK chain}
The Hamiltonian for the BBQK chain with $L$ sites and open boundary condition (OBC) is given by
\begin{eqnarray}
  {H}=&&
  J_1 \sum_{j=1}^{L-1} \bm{S}_j\cdot \bm{S}_{j+1}
  + J_2 \sum_{j=1}^{L-1} (\bm{S}_j\cdot \bm{S}_{j+1})^2
  \nonumber\\
  &&+
  K \sum_{j=1}^{\lfloor L/2 \rfloor} ( S_{2j-1}^xS_{2j}^x + S_{2j}^yS_{2j+1}^y),
\label{eq:HBBQK}
\end{eqnarray}
where the exchange parameters are normalized onto a unit sphere as
\begin{equation}
  (J_1,J_2,K)=(\sin \theta \cos \phi, \sin \theta \sin \phi,\cos \theta).
\label{eq:para}
\end{equation}
Here, $S^{\alpha}_j$ ($\alpha=x,\,y,\,z$) are spin-1 operators at site $j$ that satisfy the commutation relations $[S^{\alpha}_i,S^{\beta}_j]=i \delta_{ij} \epsilon_{\alpha\beta\gamma} S^{\gamma}_j$. In this work, we focus on the northern hemisphere ($0 \leq \theta \leq \pi/2$), corresponding to antiferromagnetic Kitaev interaction ($K \geq 0$)~\cite{PhysRevB.99.241106,Kim_2022}.

At the equator ($\theta=\pi/2$), Eq.~\eqref{eq:HBBQK} reduces to the BBQ chain, which hosts a rich phase diagram that has been studied extensively~\cite{1993TXiangBBQchain,1995BBQchain,1997Itoi,2011BBQwithZeemanOptical,2011BBQwithZeeman,2017BBQwithZeeman}. A key feature of this model is the exactly solvable Affleck–Kennedy–Lieb–Tasaki (AKLT) point at $\phi=\arctan 1/3$ ($J_2/J_1=1/3$)~\cite{1988AKLT,1992KTtransformation,Kennedy1992Z2Z2}. The AKLT ground state is a gapped, topological valence-bond solid that realizes the Haldane phase~\cite{1983Haldane,1988Haldane} and is characterized by a finite nonlocal string order parameter~\cite{denNijs1989}. The Haldane phase itself is bounded by two Bethe ansatz solvable critical points at $\phi=-\pi/4$~\cite{1989BBQspectrum} and $\pi/4$~\cite{1975MulticomponentBethe}, described by SU(2)$_2$~\cite{1987AffleckHaldane} and SU(3)$_1$~\cite{1994Itoi} Wess-Zumino-Witten (WZW) effective field theories, respectively.
For $\phi>\pi/4$, the system enters a gapless quadrupole phase with soft modes at momenta $k = \pm 2\pi/3$ emerges ~\cite{1991triplingBBQ,1997Itoi,2006BBQchain}. This phase extends to another Bethe-ansatz solvable, SU(3) symmetric point at $\phi=\pi/2$~\cite{1975MulticomponentBethe,1989BBQspectrum,Klmper1989NewRF,1990corrBiquadratic}, which marks the onset of a ferromagnetic phase. The phase diagram of the BBQ chain is completed by a dimerized phase driven by the dominant biquadratic interaction between the pair of SU(3) symmetric points at $\phi=5\pi/4$ and $7\pi/4$~\cite{2006BBQchain}.
This dimerized phase is known to be unstable to perturbations. For instance, it has been shown that quadratic Zeeman terms drive a transition into a nematic phase~\cite{2011BBQwithZeemanOptical,2011BBQwithZeeman,2017BBQwithZeeman}.
A critical nematic phase near $\phi = 5\pi/4$, conjectured in earlier studies~\cite{1991Chubukov}, has been ruled out by various numerical works~\cite{2005BasenceNematic,2005PhysRevLett.95.240404,2006RGBBQ,2006BBQchain,2017Rakov}, although $\theta\to (5\pi/4)^+$ remains asymptotically close to two nearby critical nematic phases under quadratic Zeeman perturbations~\cite{dai2023absencecriticalnematicphase}.

Introducing the Kitaev interaction shifts the system away from the equator and breaks the global spin SU(2) symmetry of the BBQ chain. Nevertheless, some discrete symmetries remain, including time reversal symmetry, $\mathcal{T}:$ $S_j^{\alpha} \mapsto -S^{\alpha}_j$,
spatial inversion symmetry, $\mathcal{I}:$ $S_j^{\alpha}\mapsto S_{N+1-j}^{\alpha}$, and global $\pi$-rotations about the $x$, $y$, and $z$ axes,  given by $\mathcal{R}_\alpha(\pi)$ = $\prod_j e^{i\pi S^\alpha_j}$.
Since only two of these global rotations are independent, the corresponding symmetry group is the dihedral group $\mathcal{D}_2=\mathbb{Z}_2 \times \mathbb{Z}_2$ ~\cite{2010Spin1KitaevTheory,2008SpinShoneycombKitaev}.
In addition, the model possesses a screw symmetry for an infinite system or with periodic boundary conditions (PBC)~\cite{WLYou2020HK}, given by the operator
\begin{equation}\label{eq:symmetry}
   \mathcal{G}=\left\{C_{4z} \vert T_1\right\}.
\end{equation}
This symmetry operator combines a global $\pi/2$ rotation around $z$-axis $C_{4z}: (S^x_j, S^y_j)\mapsto(S^y_j,-S^x_j)$, which interchanges the $x$ and $y$-Kitaev coupling directions on a given bond, with a one-site translation, $T_1$, that restores the original Hamiltonian.

Analogous to the 2D honeycomb case, a $\mathbb{Z}_2$ invariant~\cite{2010Spin1KitaevTheory,2008SpinShoneycombKitaev} can be defined at the pure Kitaev point ($\theta=0$) for the spin chain~\eqref{eq:HBBQK}.
To see this, we first define the site-parity operators $\Sigma_j^{\alpha}=e^{i\pi S_j^{\alpha}}=I-2(S^{\alpha}_j)^2$,
which satisfy $\Sigma_j^x\Sigma_j^y\Sigma^z_j=I$ with $I$ being the identity. From this, the bond parity operators are introduced as
\begin{equation}\label{Z2invariant}
  W_{2j-1}=\Sigma^y_{2j-1}\Sigma^y_{2j}, \quad W_{2j}=\Sigma^x_{2j}\Sigma^x_{2j+1}.
\end{equation}
These operators commute with each other as well as the pure Kitaev chain Hamiltonian $H_K=H(\theta=0)$, serving as $\mathbb{Z}_2$ invariants of $H_K$ with eigenvalues $\pm 1$. The full Hilbert space of $H_K$ is partitioned into sectors labeled by the $\mathbb{Z}_2$ values $w_j=\pm 1$. The ground state of the pure Kitaev chain lies in the flux-free sector, where all $w_j = +1$~\cite{2010Spin1KitaevTheory}.

\section{\label{sec:methods}Method}
As the generic BBQK model is not exactly solvable, we determine its ground state and measure observables using the finite DMRG method within the matrix-product-state (MPS) framework. To analyze the results and identify possible phases, we evaluate various order parameters and correlation functions relevant to specific phases. Since spontaneous symmetry breaking does not occur in finite systems, a pinning field is occasionally applied at the ends of the chain to select a specific symmetry sector when degeneracy occurs~\cite{Stoudenmire2012review}. In such cases, the effects of pinning are carefully assessed, and the robustness of our conclusions is rigorously verified.

In addition, quantum-information inspired tools offer valuable insights into quantum phases and their transitions. A widely studied quantity in this context is the entanglement entropy, which serves as a key diagnostic for quantum criticality~\cite{AmicoRMP2008}. It is defined as
\begin{equation}\label{eq:ee}
  S= -\Tr\rho_A \ln \rho_A,
\end{equation}
where $\rho_A$ is the reduced density matrix of a subsystem $A$ in a bipartition $A\cup B$, obtained by tracing out its complement $B$. Another useful quantity is the entanglement spectrum, defined as the eigenvalue spectrum of $\rho_A$, which has proven particularly powerful in identifying topologically nontrivial phases, such as fractional quantum Hall states~\cite{LiHaldane2008} and the Haldane phase~\cite{Pollmann2010ESofSPT}. Both the entanglement entropy and entanglement spectrum can be directly extracted from the Schmidt coefficients within MPS-DMRG calculations. In 1D gapped spin systems with local Hamiltonians, the entanglement entropy $S$ generally satisfies area law~\cite{2010AreaLaws} and thus remains bounded.
At critical points, however, $S$ diverges logarithmically with a prefactor determined by the central charge of the underline conformal field theory (CFT)~\cite{2010AreaLaws,Calabrese2009CFT} that aids in determining the universality and critical exponents of the phase transition. Comparing numerical results to known CFT predictions at critical points also helps assess computational convergence and accuracy.

Fidelity measures similarity between two quantum states by measuring their overlap $F = |\langle \psi|\phi \rangle |$. It is naturally expected to show sensitivity near critical points, where the ground state structure changes non-analytically~\cite{Paolo2006Overlap}. To capture this behavior more systematically, one can define the fidelity susceptibility~\cite{WLYou2007FideSusc,2010FidelityQPT,2019FidelityDQCP}
\begin{equation}
  \chi_F=-2\lim_{\delta \lambda \to 0} \frac{\ln (|\langle \psi_0(\lambda)|\psi_0(\lambda+\delta \lambda)\rangle|)}{(\delta \lambda )^2},
\end{equation}
which measures the leading response of the ground-state fidelity to a small change in a Hamiltonian parameter $\lambda$. At quantum critical points, $\chi_F$ typically exhibits pronounced peaks that diverge in the thermodynamic limit. In finite systems, scaling relations between $\chi_F$ and critical exponents can yield further insights into the nature and universality class of the transition~\cite{QMC2009Fidelity,Albuquerque2010QCriticalScaling}.

\section{\label{sec:result}Results}

The phase diagram of the BBQK chain for $K \geq 0$ is shown in Fig.~\ref{Fig_PD}, based on results from finite chains of up to $L=512$ sites. MPS bond dimensions are kept up to $\chi = 1000$, ensuring minimal truncation errors, typically $\epsilon < 10^{-10}$ for gapless phases and $\epsilon < 10^{-14}$ for gapped phases. All calculations are performed with OBC unless otherwise specified.

Before discussing the detailed characterization of individual phases, we first provide an overview of the phase diagram in Fig.~\ref{Fig_PD}. As discussed in Sec.~\ref{sec:model}, the BBQ model at the equator ($\theta = \pi/2$) exhibits four distinct phases. Upon introducing the Kitaev interaction, most of the phases remain stable over a sizable range of $\theta$, except for the dimerized phase induced by the dominant biquadratic interaction in the region $\phi \in (5\pi/4,\,7\pi/4$), which transitions immediately into a Kitaev nematic phase and will be discussed in detail in Secs.~\ref{sec:result_A} and~\ref{sec:result_B}. This phase encloses the pure Kitaev point, which was previously identified as a QSL phase due to its vanishing spin-spin correlation beyond nearest neighbors~\cite{WLYou2020HK,2022WLYouManybodyscar}.
In contrast, starting from the gapless quadrupole phase [$\phi \in (\pi/4, \pi/2)$], increasing $K$ induces another dimerized phase, which we refer to as the Kitaev dimer phase. This phase spontaneously breaks the $x$-$y$ symmetry and develops a finite dimer order in the $x$ or $y$ spin components in the thermodynamic limit, which will be detailed in Sec.~\ref{sec:result_C}. These two dimer phases arise from the interplay between the BBQ and Kitaev interactions and are absent in either model individually.

Moving away from the pure Kitaev point while keeping $\phi = 0$ or $\phi = \pi$ introduces the antiferromagnetic or ferromagnetic bilinear couplings $J_1$, respectively, while keeping the biquadratic coupling $J_2 = 0$. On the antiferromagnetic side, the system transitions rapidly into the gapped Haldane phase, whereas on the ferromagnetic side, the system enters a topological left-left-right-right (LLRR) phase, where the dominant two-point spin correlation functions exhibit a four-site $(+,+,-,-)$ pattern, before eventually transitioning into the fully polarized ferromagnetic phase. These findings are consistent with previous studies on $J_1$-$K$ spin-1 chains~\cite{WLYou2020HK,2022WLYouManybodyscar,2023HKSIA}.

In the following, we discuss the phase diagram in detail, focusing on phases intersected by the red dashed lines in Fig.~\ref{Fig_PD}, including the newly identified Kitaev nematic and Kitaev dimer phases. The ferromagnetic, LLRR, and the Haldane phases have been studied extensively in previous works~\cite{1983Haldane,1988Haldane,WLYou2020HK}, and will not be the focus here. For completeness, their main characteristics are summarized in Appendix~\ref{app:1}.

\subsection{\label{sec:result_A}Biquadratic dimer to Kitaev nematic phase transition}

We begin at the pure biquadratic point $(\theta, \phi)=(0.5\pi, 1.5\pi)$, where the biquadratic exchange $J_2=-1$ remains the only non-vanishing interaction. This point lies at the center of the biquadratic dimer phase that breaks one-site translation symmetry. This phase is characterized by a local dimer order parameter
\begin{equation}\label{eq:DimerOrderParameter}
  D^{\alpha\alpha}(r)=\langle S_{r-1}^{\alpha}S^{\alpha}_r-S^{\alpha}_rS_{r+1}^{\alpha}\rangle,
\end{equation}
which quantifies the asymmetry in spin-spin correlations on adjacent bonds at site $r$ for spin components $\alpha \in \{x,\,y,\,z\}$. The bond-summed total dimer order parameter is then defined as
\begin{equation}\label{eq:ToalDimer}
D_\mathrm{tot}=\biggl|\sum_{\alpha}D^{\alpha\alpha} \biggr|
  =\biggl|
\langle \bm{S}_{r-1}\cdot \bm{S}_r-\bm{S}_r\cdot \bm{S}_{r+1}\rangle\biggr|.
\end{equation}
Previous studies have shown that this phase possesses a small excitation gap, $\Delta\approx 0.17|J_2|$, and a long correlation length, $\xi\approx 21$ ~\cite{1989BBQspectrum, Klmper1989NewRF,1990corrBiquadratic}.

\begin{figure}[t]
  \includegraphics[width=\linewidth]{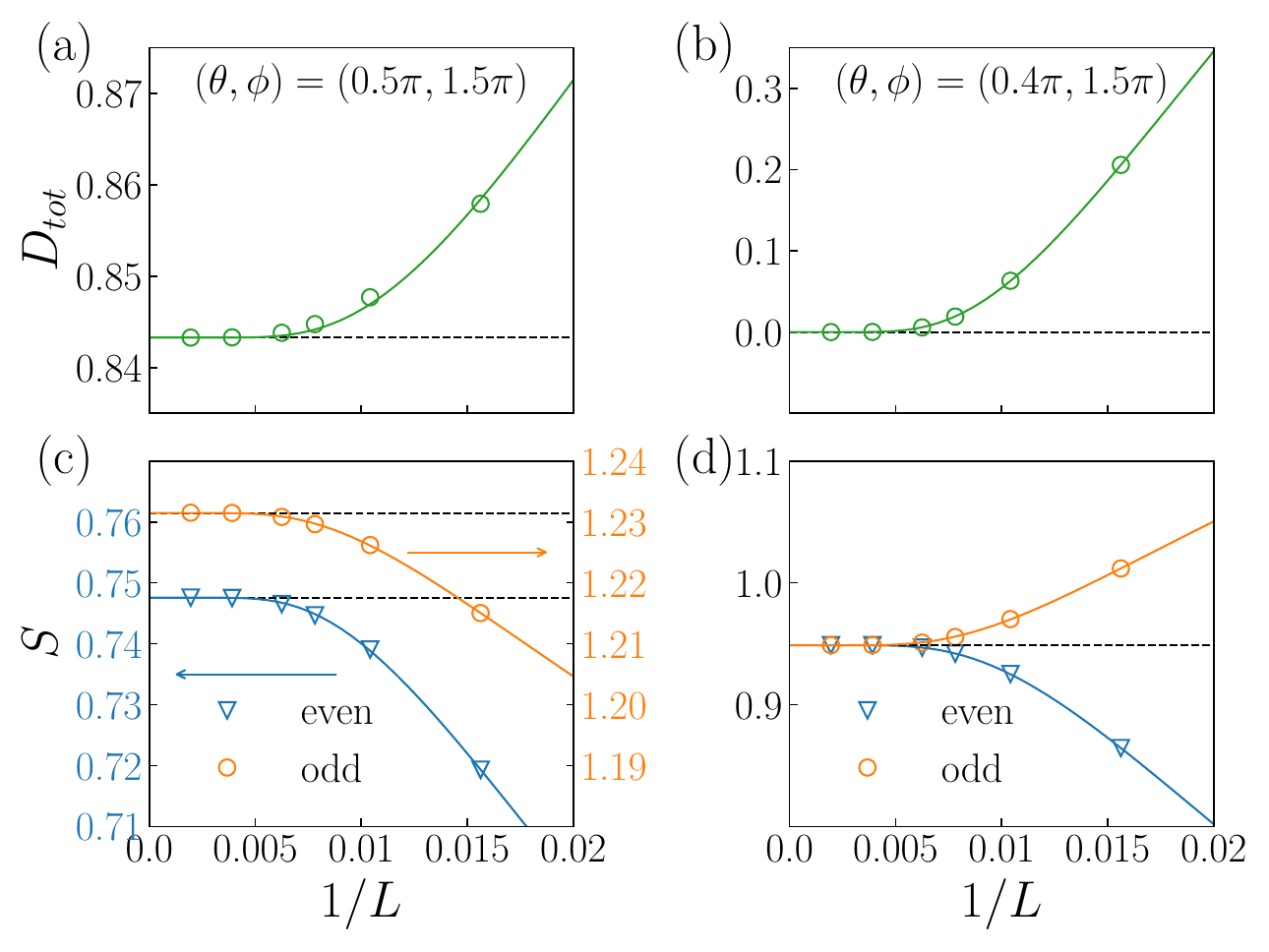}
  \caption{\label{fig:DimerDis}
(a)-(b) Total dimer order parameter $D_\mathrm{tot}$ and (c)-(d) entanglement entropy $S$ at even and odd sites for $(\theta, \phi) = (0.5\pi, 1.5\pi)$ in (a), (c) and $(0.4\pi, 1.5\pi)$ in (b), (d). Here, $D_\mathrm{tot}$ is averaged over the 32 central sites to capture bulk behavior. Results are shown for system sizes $L = 64$, 96, 128, 160, 256, and 512. Solid lines in each panel represent extrapolations to the thermodynamic limit (see text).
}
\end{figure}

Figure~\ref{fig:DimerDis}(a) shows the calculated $D_\mathrm{tot}$ at the biquadratic point for different chain lengths, demonstrating convergence to a finite value as $L$ increases. Extrapolation using the form $D_\mathrm{tot}(L) = D_\mathrm{tot}^{\infty} + c_1 e^{-c_2 L}$~\cite{1995BBQchain} yields $D_\mathrm{tot}^{\infty} = 0.8434$ in the thermodynamic limit, with a fitted correlation length $\xi = 1/c_2 \approx 22.51$, in good agreement with the Bethe ansatz result $\xi \approx 21$ ~\cite{1989BBQspectrum, Klmper1989NewRF,1990corrBiquadratic}. Here and throughout the paper, $c_1$ and $c_2$ denote real-valued fitting parameters.
This finite dimer order is also reflected in the entanglement structure. With PBC, the ground state is twofold degenerate, with stronger bonding on either even or odd bonds. With OBC, however, edge spins are unpaired on one side and can only form bonds with adjacent spins on the other. For chains of even length, this asymmetry favors dimerization on odd bonds.
This behavior is evident in the entanglement entropy shown in Fig.~\ref{fig:DimerDis}(c). For the same system size $L$, $S_\mathrm{odd}$ is consistently larger than $S_\mathrm{even}$, reflecting the stronger dimerization on odd bonds induced by OBC. As $L$ increases, both quantities converge to constant values, consistent with the area law for entanglement entropy in gapped one-dimensional systems. Using the extrapolation form $S(L) = S^{\infty} + c_1 e^{-c_2L}$, the thermodynamic values are estimated to be 0.748 and 1.231 for $S_\mathrm{even}$ and $S_\mathrm{odd}$, respectively.

Moving upward from the biquadratic point in Fig.~\ref{Fig_PD} introduces a finite Kitaev interaction $K$, which is expected to disrupt the dimer order due to its bond-dependent nature. Figure~\ref{fig:DimerDis}(b) shows the calculated $D_\mathrm{tot}$ at $(\theta,\phi)=(0.4\pi, 1.5\pi)$,  which decreases with increasing $L$ and extrapolates to zero within an error of $10^{-6}$ as $L \to \infty$, indicating the absence of dimerization in the thermodynamic limit. This conclusion is further supported by the entanglement entropy in Fig.~\ref{fig:DimerDis}(d), where both $S_\mathrm{even}$ and $S_\mathrm{odd}$ converge to the same value 0.9488 within an error of $10^{-5}$. The point $(0.4\pi, 1.5\pi)$ thus lies within a distinct, non-dimerized phase that remains to be characterized.

\begin{figure}[tb]
  \includegraphics[width=\linewidth]{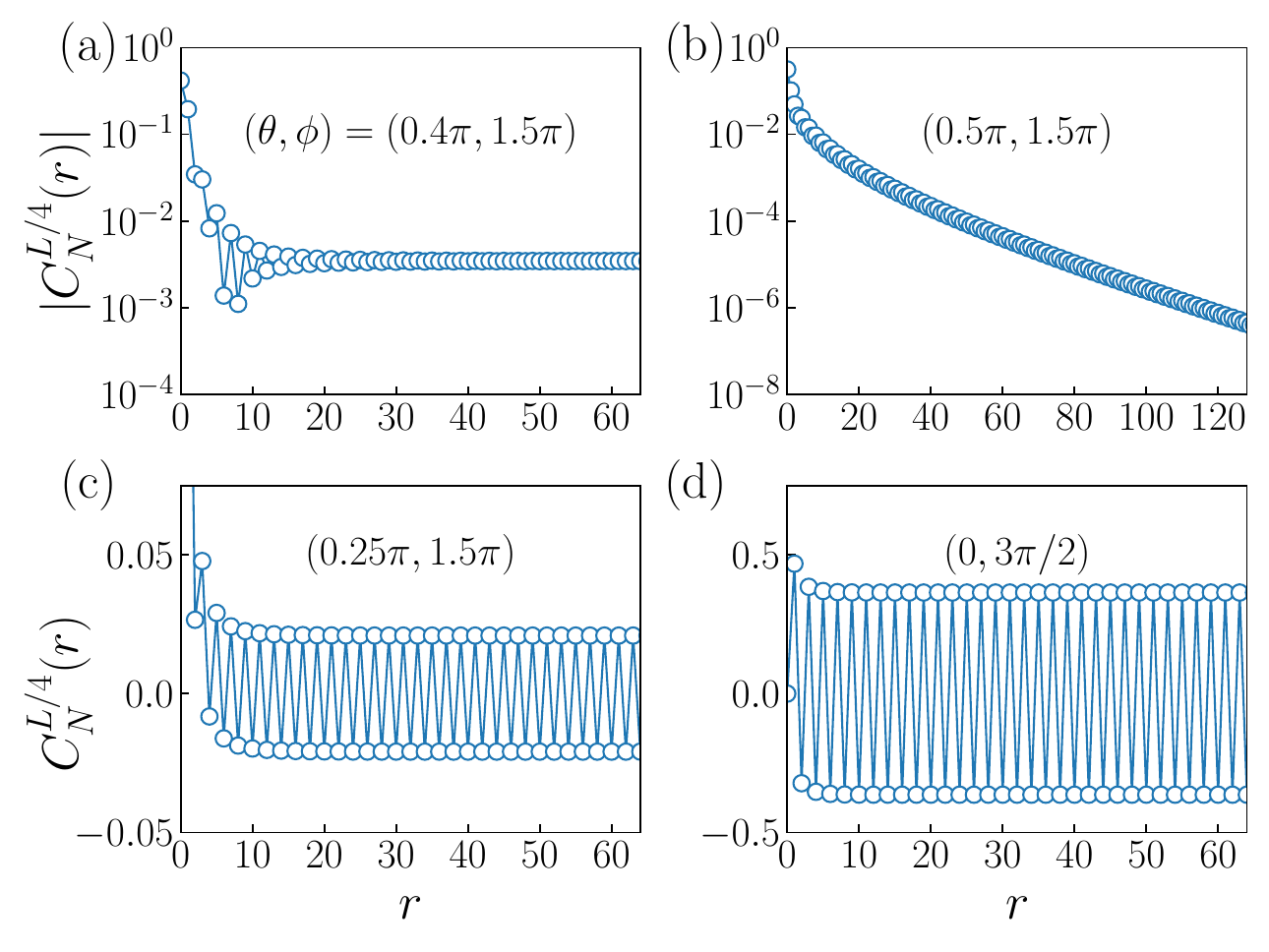}
  \caption{\label{fig:NematicCorr}
  Nematic correlation functions $C_N^{L/4}(r)=C_N(L/4,L/4+r)$ at (a) $(\theta, \phi)=(0.4\pi, 1.5\pi)$, (b) $(0.5\pi, 1.5\pi)$, (c) $(0.25\pi,1.5\pi)$ and (d) Kitaev point $(0, 1.5\pi)$, calculated for $L=512$.
}
\end{figure}

Previous studies suggest that the dimer order is unstable under perturbations and may give way to a nematic phase~\cite{2011BBQwithZeemanOptical,2011BBQwithZeeman,2017BBQwithZeeman}. To investigate the possible nematic nature of the phase surrounding $(0.4\pi, 1.5\pi)$, we compute the four-spin correlation function~\cite{2012Syromyatnikov,2013PhysRevLett.110.077206,Luo2023nematic}
\begin{equation}
\label{eq:nematic_ord}
C_N(i,j) = \langle S^+_i S^+_{i+1} S^-_j S^-_{j+1}\rangle \sim O_N^2 e^{i\varphi},
\end{equation}
from which the spin nematic order parameter $O_N$ can be extracted. Here, $\varphi$ is an interaction-dependent phase factor that is a priori unknown. For the $J_2$-$K$ Hamiltonian, we find that $C_N(i,j)$ is real-valued with alternating signs between even and odd bond distance $|i-j|$, corresponding to $\varphi = \pi$. We therefore define $C^{i}_N(r) = |C_N(i, i+r)|$ and plot its behavior across the bulk of the chain in Figs.~\ref{fig:NematicCorr}(a) and (b) for $(0.4\pi, 1.5\pi)$ and $(0.5\pi, 1.5\pi)$, respectively. In the former case, $C^{L/4}_N(r)$ initially decays with $r$ but quickly converges to a finite value, indicating the presence of nematic order. We refer to the surrounding phase as the Kitaev nematic phase, as the nematic behavior arises directly from the underlying Kitaev interaction. In contrast, it decays exponentially in the latter case, signaling the absence of nematic correlations.

\begin{figure}[tb]
  \includegraphics[width=\linewidth]{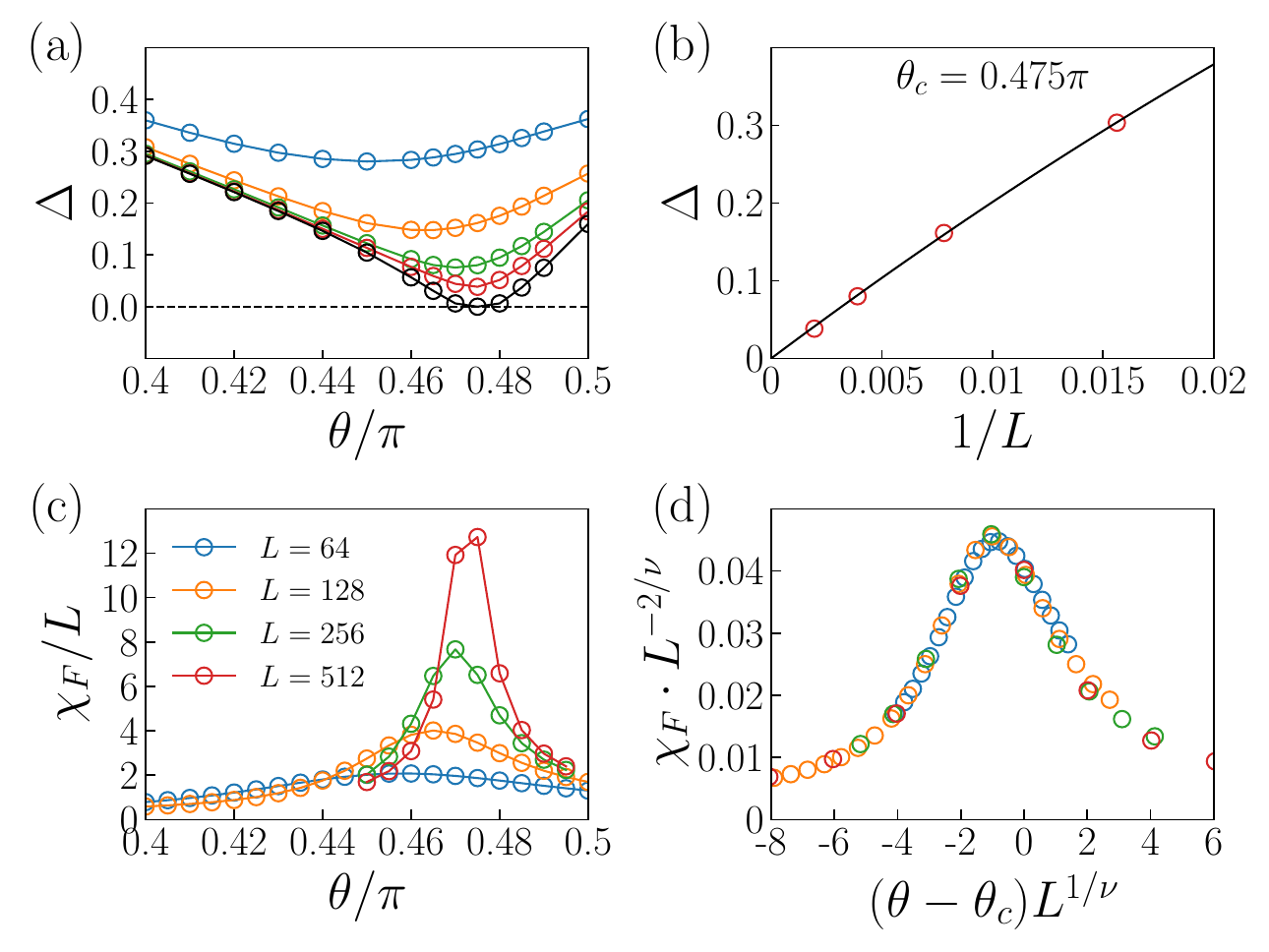}
  \caption{\label{fig:GapClose}
  (a) Excitation gap $\Delta$ for $L=64$, $128$, $256$, and $512$ with $0.4\pi < \theta < 0.5\pi$ and $\phi = 1.5\pi$. The extrapolated values $\Delta_{\infty}$ are shown in black.
  (b) Finite-size scaling of the gap size at $\theta_c=0.475\pi$.
  (c) Per-site fidelity susceptibility $\chi_F(L)/L$.
  (d) Data collapses of fidelity susceptibility via Eq.~(\ref{eq:data_collapse}).
}
\end{figure}

Having established the distinct nature of the biquadratic dimer and Kitaev nematic phases, we now turn to the precise determination of the quantum critical point separating them. Since both phases are gapped, the transition between them must be accompanied by a closing of the excitation gap. To locate this transition, we compute the gap $\Delta = E_1 - E_0$ between the first excited state $E_1$ and the ground state $E_0$ at fixed $\phi = 1.5\pi$ as a function of $\theta$. Figure~\ref{fig:GapClose}(a) shows the calculated $\Delta$ for various system sizes $L$, revealing a significant suppression of the gap within the range $0.4\pi < \theta < 0.5\pi$, with the magnitude and precise location of the minimum depending on $L$. Finite-size scaling using the form $\Delta(L)=\Delta_{\infty}+c_1/L+c_2/L^2$ indicates that the gap closes at the critical point $\theta_c \approx 0.475\pi$ [Fig.~\ref{fig:GapClose}(b)], corresponding to a coupling ratio $K/J_2 \approx -0.0787$. The narrow stability range of the dimer phase aligns with previous studies and further corroborates its fragility against perturbations~\cite{2011BBQwithZeemanOptical,2011BBQwithZeeman,2017BBQwithZeeman}.

To further investigate the nature of the phase transition and confirm $\theta_c$, we compute the fidelity susceptibility $\chi_F$ near the transition, as shown in Fig.~\ref{fig:GapClose}(c). The per-site fidelity susceptibility $\chi_F/L$ exhibits a peak that sharpens and shifts toward $\theta_c$ as $L$ increases. This allows us to perform finite-size analysis using the relation~\cite{QMC2009Fidelity,Albuquerque2010QCriticalScaling}
\begin{equation}
\label{eq:data_collapse}
L^{-d} \chi_F(\theta,L)= L^{2/\nu-d} f_{\chi_F}(L^{1/\nu}|\theta-\theta_c|),
\end{equation}
where $d = 1$ is the system dimension and $\nu$ is the critical exponent governing the divergence of the correlation length $\xi$, $\xi \sim |\theta - \theta_c|^{-\nu}$ near the critical point. The function $f_{\chi_F}$ is a homogeneous scaling function yet to be determined. Figure~\ref{fig:GapClose}(d) shows the data collapse of the per-site fidelity susceptibility according Eq.~\eqref{eq:data_collapse}. From this collapse, we extract a critical point $\theta_c \approx 0.475\pi$, in excellent agreement with the value obtained from the gap analysis above. The fitted critical exponent is $\nu \approx 1.01$, consistent with the Ising universality class value $\nu_\mathrm{Ising} = 1$.

\begin{figure}[t]
    \includegraphics[width=\linewidth]{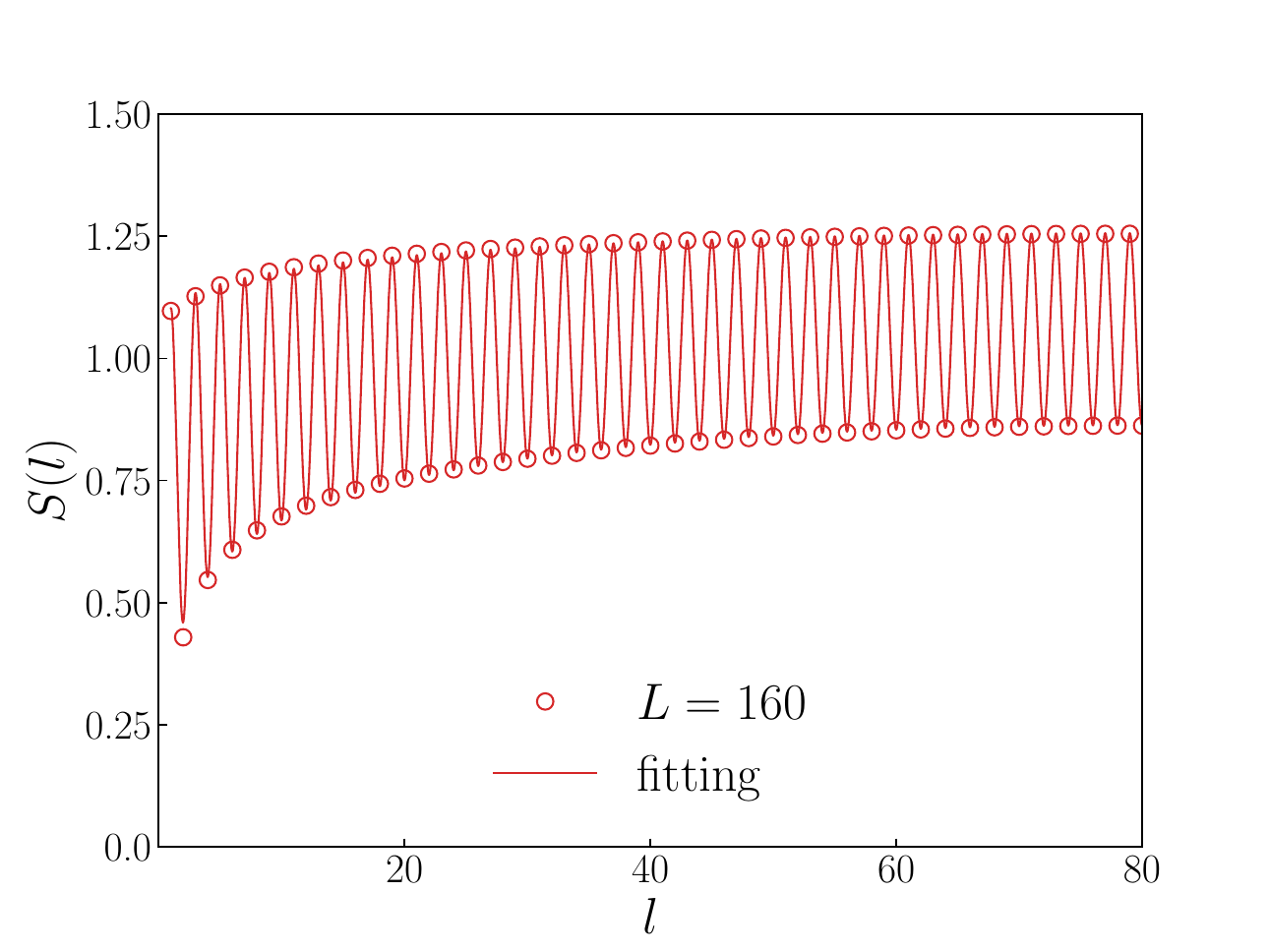}
    \caption{\label{fig:centralcharge}
    Entanglement entropy over half of the chain with $L=160$ at $(\theta_c,\phi)=(0.475\pi,1.5\pi)$. The red line shows the fit with $p=0.160$ and $c=0.499$ using Eq.~\eqref{eq:centralcharge}.
   }
 \end{figure}

Additional evidence for Ising universality can be obtained from the entanglement behavior near $\theta_c$. Figure~\ref{fig:centralcharge} shows the entanglement entropy $S(l)$ between the leftmost $l$ sites and the remaining $L - l$ sites for $L=160$, exhibiting pronounced even-odd oscillations. Inspired by previous studies on spin chains~\cite{2010Calabrese,2015SUNRenyiEntropy}, we fit $S(l)$ to the scaling form
\begin{equation}\label{eq:centralcharge}
    S(l)=S_\mathrm{log} + S_\mathrm{osc} + \tilde{c},
\end{equation}
where $S_\mathrm{log}$ is of the standard CFT prediction for a finite chain with OBC~\cite{Calabrese2009CFT}
\begin{equation}\label{eq:centralcharge_log}
    S_\mathrm{log}= \frac{c}{6} \ln \biggl[\frac{2L}{\pi}\sin(\pi\frac{l}{L})\biggr],
\end{equation}
and $S_\mathrm{osc}$ is an oscillatory correction of the form
\begin{equation}
    S_\mathrm{osc}(l) = F(\frac{l}{L}) \cos(2k_Fl)\biggl|
    \frac{4L}{\pi} \sin(\pi\frac{l}{L})
    \biggr|^{-p},
\end{equation}
with $\tilde{c}$ being a non-universal constant. Here, $k_F=\pi/2$ is a characteristic ``Fermi momentum'' chosen to capture the observed even-odd oscillations. The function $F(l/L)$ is a universal scaling function, which we approximate as constant to facilitate fitting. The critical exponent $p$, which formally encodes information about the underlying CFT~\cite{2010Calabrese,2015SUNRenyiEntropy}, is treated as a free fitting parameter under this approximation. The best fit yields a central charge $c = 0.499$ and exponent $p=0.160$, in excellent agreement with the Ising universality class value $c_\mathrm{Ising}= 1/2$.

\subsection{\label{sec:result_B}Kitaev nematic phase and the pure Kitaev point}

The phase surrounding the pure Kitaev point ($\theta = 0$) is characterized by vanishing two-point spin-spin correlations beyond nearest neighbors and was therefore previously identified as the Kitaev QSL phase in Refs.~\cite{WLYou2020HK,2022WLYouManybodyscar}. However, it has also been shown that this phase exhibits nonvanishing nematic order~\cite{Luo2023nematic,PhysRevResearch.3.033048}, similar to the Kitaev nematic phase. This shared nematicity motivates a closer examination of their connection and distinction.

\begin{figure}[t]
    \includegraphics[width=\linewidth]{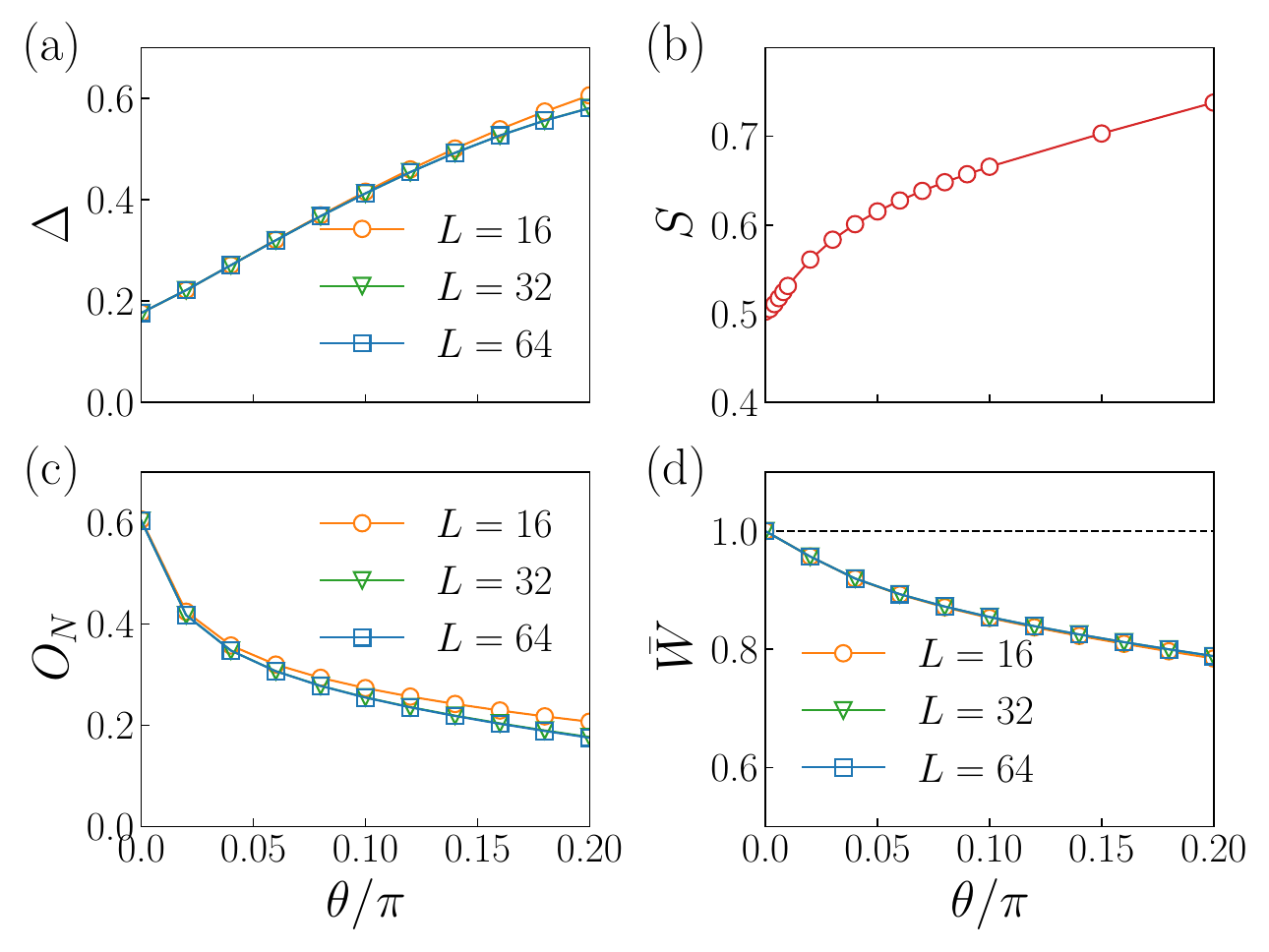}
    \caption{\label{fig:crossover}
    (a) Energy gap $\Delta$, (b) half-chain entanglement entropy $S(L/2)$, (c) nematic order parameter $O_N$, and (d) averaged bond operator $\bar W$ as a function of $\theta$, computed at fixed $\phi = 1.5\pi$. Panels (a), (c), and (d) are calculated for $L = 16$, $32$, and $64$ with PBC, while panel (b) shows results for $L = 65$ with OBC. An odd chain length under OBC ensures equal numbers of even and odd bonds.
   }
\end{figure}

Figure~\ref{fig:crossover}(a) shows the excitation gap $\Delta$ as a function of $\theta$ at fixed $\phi = 1.5\pi$ calculated with PBC to eliminate edge effects of the Kitaev term~\cite{Luo2023nematic}. At $\theta = 0$, the gap is $\Delta = 0.176$, consistent with previous exact diagonalization estimates, which place it within the range $0.1671 < \Delta < 0.1802$~\cite{2010Spin1KitaevTheory}. With increasing $\theta$, the excitation gap increases monotonically and shows no sign of reduction or closing, indicating the absence of a phase transition. This conclusion is further supported by the smooth behavior of the half-chain entanglement entropy $S(L/2)$ shown in Fig.~\ref{fig:crossover}(b), which exhibits no anomalies across the entire range. The nematic order parameter $O_N$, shown in Fig.~\ref{fig:crossover}(c), also remains finite throughout, as expected. Finally, we examine the behavior of the bond-parity operator. At the pure Kitaev point, as discussed earlier, the ground state is unique and characterized by a uniform bond-parity expectation value $\langle W_j \rangle = 1$ for all $j$. The biquadratic $J_2$ term does not commute with either the Kitaev interaction or the bond-parity operators, and is therefore expected to perturb this structure. Figure~\ref{fig:crossover}(d) shows the evolution of the averaged bond-parity density $\bar{W} = 1/L \sum_{j=1}^{L} \langle W_j \rangle$ with PBC, which varies smoothly without abrupt changes.

\begin{figure}[t]
    \includegraphics[width=\linewidth]{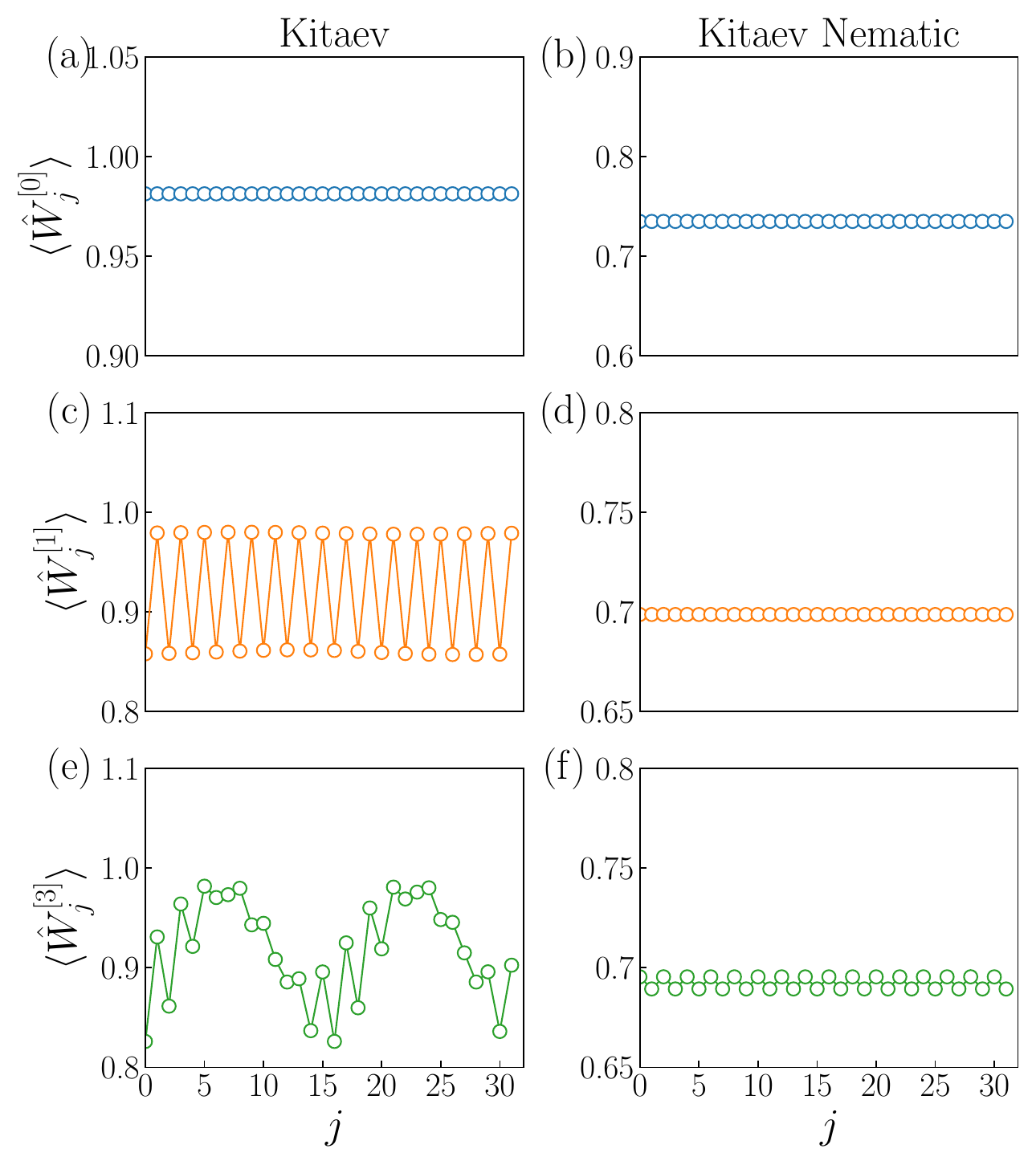}
    \caption{\label{fig:crossoverz2}
    Expectation values of the bond parity operators for $l$-th lowest state $\langle \hat{W}^{[l]}_j\rangle$ with PBC.
    Panels (a), (c), and (e) correspond to $(\theta, \phi) = (0.01\pi, 1.5\pi)$, and panels (b), (d), and (f) correspond to $(\theta, \phi) = (0.3\pi, 1.5\pi)$.
   }
\end{figure}

A previous study investigated the nature of nematic order in the spin-1 Kitaev chain with single-ion anisotropy by analyzing the spatial structure of low-lying excited states~\cite{Luo2023nematic}. There, the authors introduced a classification of distinct spin-nematic phases based on the real-space periodicity $p$ of the bond-parity expectation values $\langle W_j \rangle$ in the first few excited states and defined the change between them as a phase crossover. A similar analysis can be applied in our setting, and the corresponding results, obtained at $(\theta, \phi) = (0.01\pi, 1.5\pi)$ near the pure Kitaev point and deep in the Kitaev nematic phase at $(0.3\pi, 1.5\pi)$ for $L=32$ with PBC, are summarized in Fig.~\ref{fig:crossoverz2}.

Figures~\ref{fig:crossoverz2}(a) and (b) show the ground-state bond-parity distributions in the Kitaev and Kitaev nematic phases, respectively. In both cases, $\langle W_j \rangle$ remains uniform across the chain, preserving a period $p = 1$, similar to the behavior at the pure Kitaev point. At the pure Kitaev point, the first excited states correspond to configurations with a single bond defect~\cite{2022WLYouManybodyscar,Luo2023nematic}, where one eigenvalue of the bond operator $W_j$ is flipped from $+1$ to $-1$. These lowest excitations form an $L$-fold degenerate manifold for a chain of length $L$ with PBC.
Since the biquadratic term does not commute with either the Kitaev interaction or the bond-parity operators, introducing a finite $J_2$ couples the single-defect states and lifts their degeneracy. Exact diagonalization of small systems reveals that the biquadratic term selectively couples configurations with defects on either all even or all odd bonds, effectively partitioning the original $L$-fold manifold into two symmetry-related subspaces. Within each subspace, the eigenstates are structurally identical, with the lowest-lying states formed as equal superpositions of single-defect configurations on either all even or all odd bonds, thereby reducing the degeneracy from $L$ to $2$.

Figure~\ref{fig:crossoverz2}(c) shows the bond-parity profile $\langle W_j \rangle$ for one of these two lowest-lying eigenstates, exhibiting a period $p = 2$. The other degenerate state displays the same structure but with the values on even and odd bonds interchanged. Higher excited states within the original $L$-fold manifold involve more intricate superpositions of the single-defect states. For instance, Fig.~\ref{fig:crossoverz2}(e) shows the third excited state, characterized by a longer modulation period $p = N/2$.
As $J_2$ increases further, the excited-state manifold may begin to mix with higher-energy states involving multiple defects, leading to a qualitatively different eigenstructure. In the Kitaev nematic phase, the excitation spectrum exhibits a different organization. The first excited state is non-degenerate and retains period $p = 1$, while the second and third excited states form a nearly degenerate subspace with period $p = 2$, as shown in Figs.~\ref{fig:crossoverz2}(d) and (f), respectively.

Based on the above results, one might be tempted to interpret the observed changes in excitation structure as indicative of a phase crossover. However, it is important to note that the precise point at which the eigenstructure changes is system-size dependent, and limitations of DMRG prevent reliable access to higher excited states to enable a more systematic comparison of the eigen spectrum. Given that the excitation gap remains finite throughout the interpolation between the Kitaev QSL and Kitaev nematic phases, we argue that these two regimes should be understood as belonging to the same phase. The pure Kitaev point, while exhibiting a distinct excitation spectrum, appears as a singular point within a broader, gapped Kitaev nematic phase.

\subsection{\label{sec:result_C}Kitaev nematic, Kitaev dimer, and \\gapless quadrupole phase transitions}

\begin{figure}[t]
  \includegraphics[width=\linewidth]{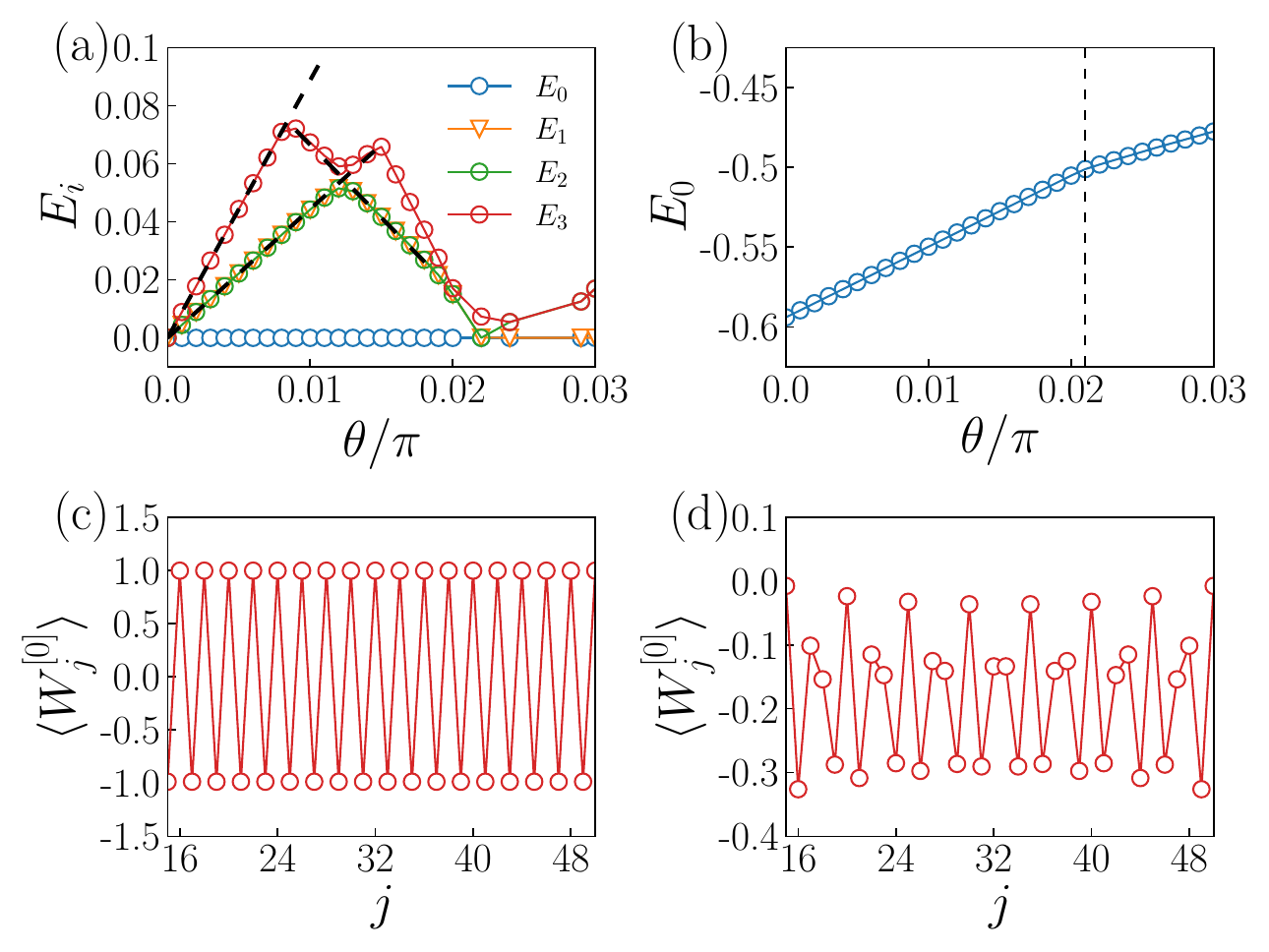}
  \caption{\label{fig:Energy_levels}
  (a) Lowest four energy eigenvalues, plotted relative to the ground-state energy in (b), as functions of $\theta$ along the path $(\theta, \phi = 0.3\pi)$, for $L = 33$ with OBC. Dashed lines in (a) are guides to the eye.
  (c,d) Ground-state expectation values of bond-parity operator $\langle W_j^{[0]}\rangle$ for $L=65$ with OBC at (c) $(\theta,\phi)=(0.1\pi,0.3\pi)$ and (d) $(0.2\pi,0.3\pi)$. }
\end{figure}

Having established the nematic nature of the phase surrounding the Kitaev point, we now proceed to explore the second segment of the red dashed line in Fig.~\ref{Fig_PD}, starting from the Kitaev point and varying $\theta$ while keeping $\phi = 0.3\pi$ fixed. Along this path, the coupling ratio remains constant at $J_2/J_1 = \tan\,(0.3\pi)\approx 1.377$, which eventually leads to the gapless phase in the BBQ chain in the absence of Kitaev coupling $K$~\cite{1991triplingBBQ,1997Itoi,2006BBQchain}.

At the pure Kitaev point, the ground state of the open spin-1 chain exhibits a fourfold degeneracy, arising from two emergent spin-1/2 edge degrees of freedom. These are labeled by the eigenvalues $w_1, w_{L-1} = \pm1$ of the boundary bond parity operators $W_1$ and $W_{L-1}$, which remain unconstrained with OBC. This degeneracy is lifted upon introducing finite $J_1$ and $J_2$. For small $\theta$, the splitting can be understood perturbatively within the four-dimensional ground-state manifold of the pure Kitaev point. The bilinear term $\sum_j \bm{S}_j\cdot \bm{S}_{j+1}$ acts as a constant in this subspace and preserves the degeneracy, whereas the biquadratic term $\sum_j(\bm{S}_j\cdot \bm{S}_{j+1})^2$ induces additive, parity-dependent positive ($+$) or negative ($-$) energy shifts from each edge. This leads to a characteristic $1_{(++)} + 2_{(+-,-+)} + 1_{(--)}$ level structure, as shown in Fig.~\ref{fig:Energy_levels}(a) for $\theta < 0.007 \pi$ for a chain of length $L=33$. As $\theta$ increases further, the higher excited states of the Kitaev chain descend and eventually close the gap, leading to an exactly twofold-degenerate ground state near $\theta \approx 0.021\pi$, marked by a kink in the energy-vs-$\theta$ curve in Fig.~\ref{fig:Energy_levels}(b), signaling a qualitative change in ground-state properties.
In contrast to the uniform bond-parity expectation values ($\langle W_j^{[0]} \rangle \approx 1$) found near the pure Kitaev point [Fig.~\ref{fig:crossoverz2}(a)], this new phase realizes a $\mathbb{Z}_2$ flux crystal, characterized by a spatially alternating pattern of $\langle W_j \rangle = \pm 1$ [Fig.~\ref{fig:Energy_levels}(c)]. Remarkably, the ground state remains an eigenstate of all $W_j$ operators, despite these operators not commuting with the full Hamiltonian. This intriguing behavior breaks down as $\theta$ increases further, as shown for $\theta = 0.2$ in Fig.~\ref{fig:Energy_levels}(d), possibly indicating the transition into another phase.

\begin{figure}[t]
  \includegraphics[width=\linewidth]{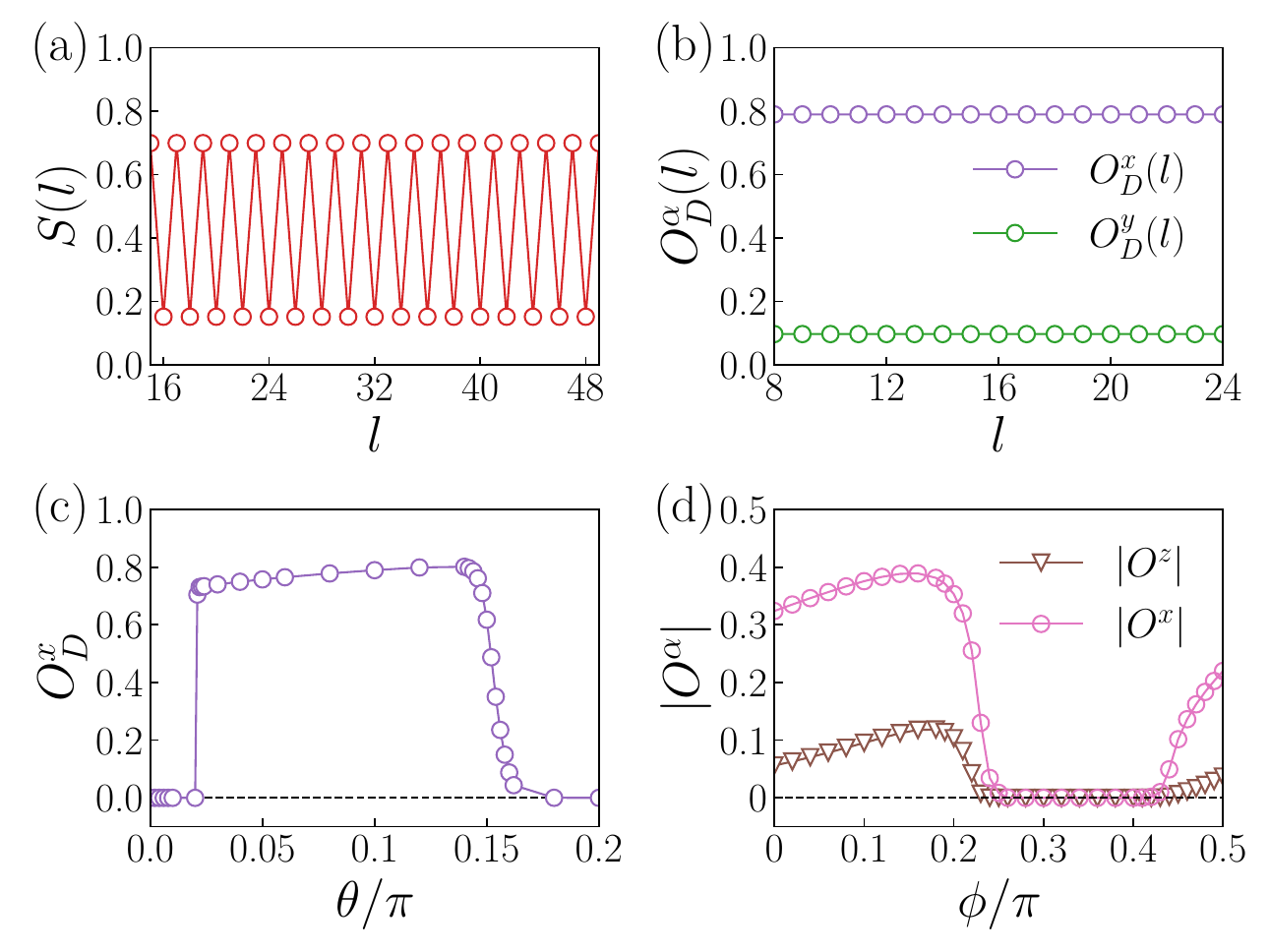}
  \caption{\label{fig:KitaevDimer}
  (a) Entanglement entropy $S(l)$ and (b) dimer order parameters $O_D^{x(y)}(l)$ at $(\theta$, $\phi)=$$(0.1\pi,0.3\pi)$ for $L=65$ with OBC,
  for one of the ground-state doublet that favors $xx$-bond.
  (c) Dimer order parameters for $0<\theta<0.2\pi$ with fixed $\phi=0.3\pi$.
  (d) String order parameters $O^{x(z)}(L/4, 3L/4)$ for $0<\phi<0.5\pi$ with fixed $\theta=0.1\pi$.
  }
\end{figure}

We first characterize the phase corresponding to Fig.~\ref{fig:Energy_levels}(c) by examining the bond-dependent entanglement entropy $S(l)$ for one of the doublet states in Fig.~\ref{fig:KitaevDimer}(a). The pronounced even-odd oscillation observed in $S(l)$ suggests the emergence of a dimer-like order. We define the following order parameters
\begin{equation}\label{eq:ODxy}
O^\alpha_D(l) = \langle S^\alpha_{l-1} S^\alpha_l - S^\beta_l S^\beta_{l+1} \rangle,
\end{equation}
where $(\alpha, \beta) = (x, y)$ or $(y, x)$. A finite value of $O^\alpha_D(l)$ signals the breaking of the screw symmetry defined in Eq.~\eqref{eq:symmetry}. Figure~\ref{fig:KitaevDimer}(b) shows the computed values of $O_D^x(l)$ and $O_D^y(l)$ in the bulk of the chain for one of the doublet states, clearly indicating the presence of finite dimer order with pronounced $x$–$y$ asymmetry. Note that $O_D^{x(y)}(l)$ are plotted only for even bonds $2j$, as the corresponding values on odd bonds related by inversion, $L + 1 - 2j$, exhibit the same amplitude but opposite sign. To confirm that this phase, referred to hereafter as the Kitaev dimer phase, is indeed distinct from the Kitaev nematic phase, we compute the dominant order parameter $O_D^x$ along the line $\phi = 3\pi/10$. As shown in Fig.~\ref{fig:KitaevDimer}(c), $O_D^x$ remains zero within the nematic phase and rises abruptly around $\theta \approx 0.02\pi$, coinciding with the point at which the gap closing and ground-state degeneracy changes in Fig.~\ref{fig:Energy_levels}(a). Upon further increasing $\theta$, $O_D^x$ drops sharply at approximately $0.16\pi$, signaling a second phase transition into the phase corresponding to Fig.~\ref{fig:Energy_levels}(d). This latter phase is identified as the gapless quadrupole phase extended from the BBQ model, as discussed below. It should be noted that, although the Kitaev dimer phase lies adjacent to both the topologically nontrivial Haldane and LLRR phases (see App.~\ref{app:1}) in Fig.~\ref{Fig_PD}, it remains topologically trivial, similar to the Kitaev nematic phase~\cite{WLYou2020HK}. This is confirmed by the behavior of the nonlocal string order parameter~\cite{1992KTtransformation,Kennedy1992Z2Z2}, defined as
\begin{equation}
\label{eq:SO}
O^{\alpha}(i,j)=\bigg\langle  S^{\alpha}_i \biggl( \prod_{k=i+1}^{j-1}\exp(i\pi S^{\alpha}_k) \biggr) S^{\alpha}_j \bigg\rangle,
\end{equation}
which serves as a standard indicator for topologically nontrivial phases. Figure~\ref{fig:KitaevDimer}(d) shows the dependence of the string order along a cut at fixed $\theta = 0.1\pi$, as $\phi$ varies across the Haldane, Kitaev dimer, and LLRR phases. The string order clearly vanishes within the dimer phase, indicating its topologically trivial nature.

Having established the Kitaev dimer phase and its main characteristics, we now turn to the transition between this phase and the gapless quadrupolar phase originally identified in the BBQ model~\cite{1991triplingBBQ,1997Itoi,2006BBQchain}. In the pure BBQ limit, the gapless quadrupolar phase is characterized by several key features. In particular, both the static spin dipole and quadrupole structure factors exhibit pronounced peaks at momentum $k_m = \pm 2\pi/3$, indicative of gapless soft modes in the excitation spectrum~\cite{2018BBQresponse,2020LowEnergyCritical}. The corresponding low-energy effective field theory is described by the SU(3)$_1$ WZW model, with central charge $c = 2$~\cite{1997Itoi}.

\begin{figure}[t]
  \includegraphics[width=\linewidth]{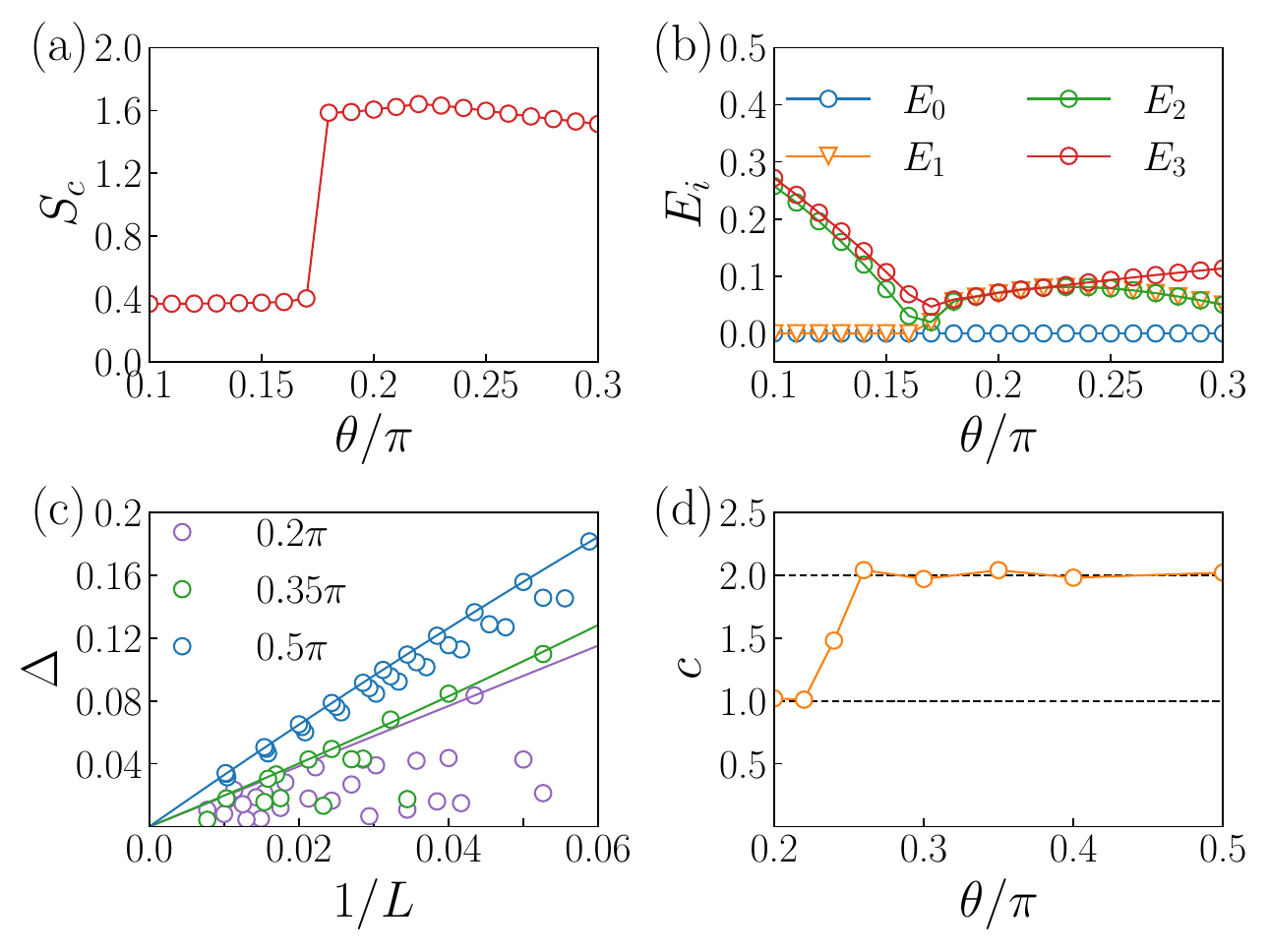}
  \caption{\label{fig:Gapless_phase}
  (a) Entanglement entropy at the center of an open chain with $L = 32$, averaged over the two central bonds as $S_c = (S_{L/2} + S_{L/2+1})/2$, plotted as a function of $\theta$ at fixed $\phi = 0.3\pi$.
  (b) Spectrum of the lowest four eigenstates for a periodic chain with $L = 10$.
  (c) Finite-size scaling of the excitation gap $\Delta$ at $\theta = 0.2\pi$, $0.35\pi$, and $0.5\pi$ with OBC. Solid lines are fits to the upper bound of the gap using $\Delta(L) = \Delta_{\infty} + c_1/L + c_2/L^2$.
  (d) Extracted central charge $c$ in the range $0.2\pi < \theta < 0.5\pi$.
  }
\end{figure}

As shown in Fig.~\ref{fig:Gapless_phase}(a), increasing $\theta$ from the Kitaev dimer phase toward the quadrupolar phase leads to a sharp rise in the entanglement entropy at the center of the chain around $0.17\pi < \theta_c < 0.18\pi$, indicating a possible phase transition. This is further supported by the spectrum of the lowest four eigenvalues in a smaller system, shown in Fig.~\ref{fig:Gapless_phase}(b), where the energy gap closes near the same critical value $\theta_c$, and the doubly degenerate ground state of the Kitaev dimer phase becomes non-degenerate.  These observations consistently identify $\theta_c$ as the critical point separating the two phases.
The finite gap observed in the nominally gapless quadrupolar phase in Fig.~\ref{fig:Gapless_phase}(b) is attributed to finite-size effects. As shown in Fig.~\ref{fig:Gapless_phase}(c), at $\theta = 0.5\pi$, where the system resides in the quadrupolar phase of the pure BBQ chain, the excitation gap $\Delta$ generally decreases with increasing system size but exhibits a pronounced three-site periodic modulation. This behavior arises from the momentum structure of the low-energy modes, as the gapless soft modes occur at $k_m = \pm 2\pi/3$, which are only commensurate with the lattice when the system length $L$ is a multiple of three. For system sizes not divisible by three, the quantized momenta do not coincide with these soft modes, resulting in a slightly larger apparent gap due to momentum mismatch. To account for this effect, we perform finite-size scaling of $\Delta$ using the largest value from each triplet, which extrapolates to zero as $L \to \infty$.
As $\theta$ is reduced, introducing a finite Kitaev interaction, this period-three structure is disrupted. Nevertheless, for $\theta = 0.35\pi$, the gap $\Delta$ still decreases with increasing $L$. A fit to the upper bound of the data still yields a vanishing gap in the thermodynamic limit. The breakdown of the periodic modulation suggests the appearance of incommensurate soft mode momenta induced by the Kitaev term. Near the phase transition at $\theta = 0.20\pi$, the gap behavior becomes more irregular but still extrapolates to zero.
Figure~\ref{fig:Gapless_phase}(d) shows the central charge $c$ extracted from fitting the half-chain entanglement entropy to $S(L/2) = c/6 \ln L + \tilde c$, omitting the oscillatory correction $S_\mathrm{osc}$ from Eq.\eqref{eq:centralcharge}, which vanishes at large $L$ and has no qualitative impact. At $\theta = 0.2\pi$, near the critical point $\theta_c$, we find $c = 1.02$. It rises rapidly to $c = 2.04$ at $\theta = 0.26\pi$, and remains at $c \approx 2.0$ within error for larger $\theta$, consistent with the gapless quadrupolar phase~\cite{2011BBQwithZeeman}. Overall, these results consistently demonstrate that the system becomes gapless for $\theta > \theta_c$, with a central charge approaching $2$, signaling the onset of a critical quadrupolar phase adiabatically connected to that of the pure BBQ model.

\begin{figure}[t]
  \includegraphics[width=\linewidth]{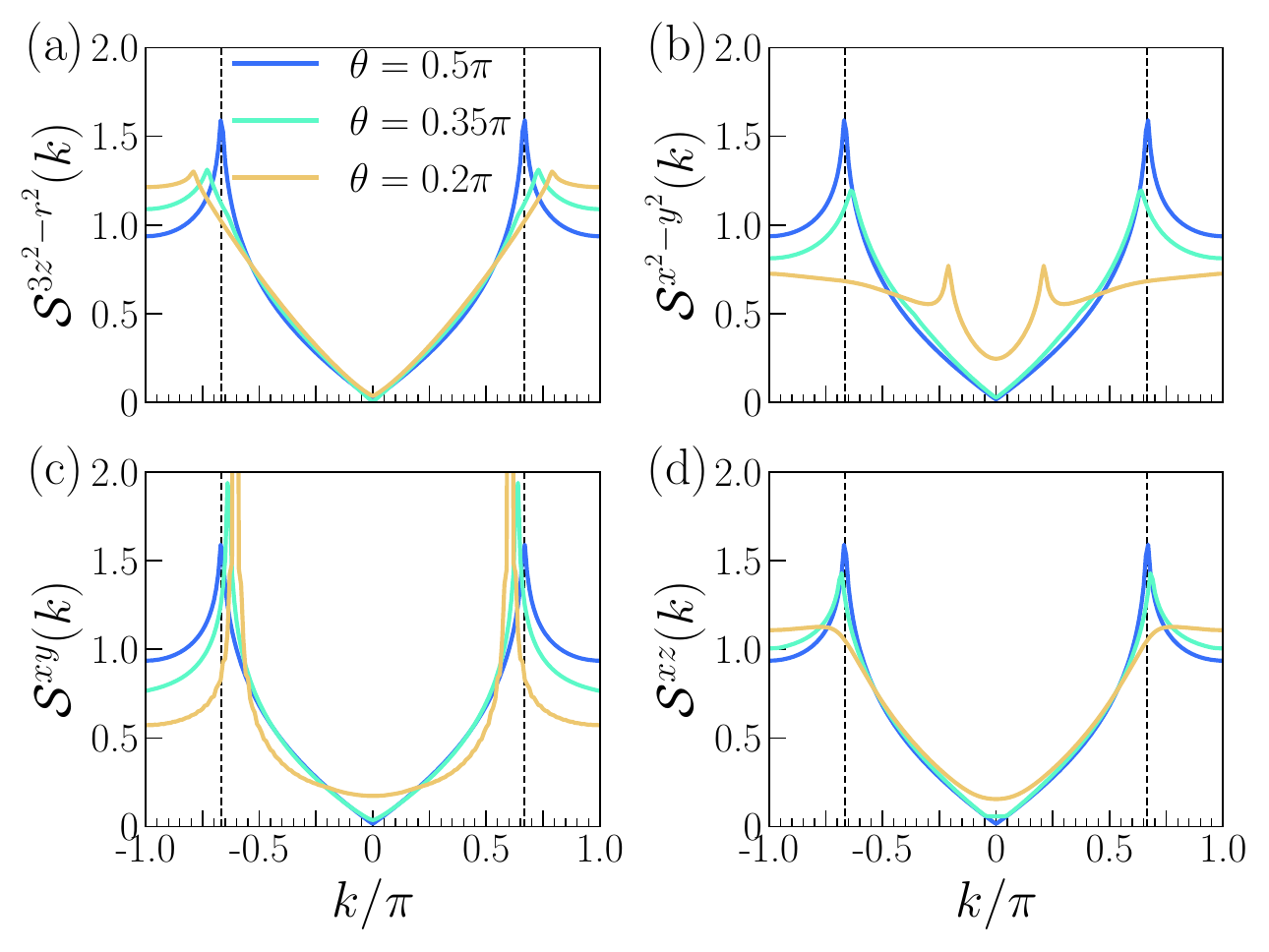}
  \caption{\label{fig:peakPosition}
    Static quadrupolar spin-spin correlations $\mathcal{S}^\mu(k)$ at $\phi = 0.3\pi$ for $\theta = 0.2\pi$, $0.35\pi$, and $0.5\pi$ on an open chain of length $L = 129$, shown for the quadrupole channels (a) $3z^2 - r^2$, (b) $x^2 - y^2$, (c) $xy$, and (d) $xz$.
  }
\end{figure}

Finally, to better understand the influence of the Kitaev interaction on the quadrupolar phase, we examine the momentum-space static quadrupolar spin-spin correlation function $\mathcal{S}^{\mu}(k)=1/L \sum_{i,j}^L \langle O^\mu_{i}O^\mu_{j} \rangle_c e^{ik(r_i-r_j)}$, where $O^\mu_i$ are the quadripole operators at site $i$
\begin{equation}\label{eq:quadrupoles}
    \mathbf{Q}_i= \begin{pmatrix}
     Q^{3z^2-r^2}_i\\Q^{x^2-y^2}_i\\Q^{xy}_i\\  Q^{xz}_i\\ Q^{yz}_i\\
    \end{pmatrix}
    =\begin{pmatrix}
        (S^z_i)^2-\frac{1}{3}S(S+1)\\
        (S^x_i)^2-(S^y_i)^2\\
        S^x_iS^y_i+S^y_iS^x_i\\
        S^x_iS^z_i+S^z_iS^x_i\\
        S^y_iS^z_i+S^z_iS^y_i\\
    \end{pmatrix},
\end{equation}
and $\langle \cdot \rangle_c$ denotes the connected correlation function.
As shown in Fig.~\ref{fig:peakPosition}, at $\theta = 0.5\pi$, the soft mode manifests as sharp peaks at $k_m = \pm 2\pi/3$ across all quadrupole channels. A finite Kitaev interaction breaks the global SU(2) symmetry of the pure BBQ model, leading to anisotropic shifts in soft-mode momenta. As $\theta$ decreases, peak positions split across channels and shift away from $\pm 2\pi/3$. In $3z^2 - r^2$ and $xz$ channels, which involve the $z$-spin component, the peaks move to higher momenta, while in the in-plane channels $x^2 - y^2$ and $xy$, they shift to lower momenta. Simultaneously, the peak intensity increases in the $xy$ channel and decreases in others, reflecting the anisotropic influence of the in-plane Kitaev term. Notably, a secondary peak emerges in the $x^2-y^2$ channel at a smaller momentum as the system approaches the phase transition [Fig.~\ref{fig:peakPosition}(b)], suggesting the possible emergence of an additional soft mode. These soft-mode momenta shifts and anisotropies shown in Fig.~\ref{fig:peakPosition} also provide a natural explanation for the irregular finite-size scaling of the excitation gap $\Delta$ observed in Fig.~\ref{fig:Gapless_phase}(c).

\section{\label{sec:conclusion}Discussion and Conclusion}
In summary, we have numerically established the ground-state phase diagram of the 1D spin-1 BBQ model with antiferromagnetic Kitaev interaction using MPS-DMRG. Our central result is the identification of two novel phases induced by the Kitaev interaction: a Kitaev nematic phase and a Kitaev dimer phase.
The former emerges from the fragile biquadratic dimer phase via a quantum critical point in the Ising universality class ($c=1/2$). We provide evidence that this gapped phase extends continuously to the pure Kitaev point, suggesting that the previously proposed Kitaev quantum spin liquid is in fact a singular limit of this broader phase, rather than a distinct one.
The gapped Kitaev dimer phase breaks the screw symmetry $\{C_{4z}|T_1\}$ with preferred $x$- or $y$- bonding, forming a state that coexists with a crystalline order of alternating $\mathbb{Z}_2$ fluxes.

Our work also sheds light on the fate of the gapless quadrupolar phase of the 1D BBQ model under Kitaev perturbations. We find that this critical phase remains stable over a broad parameter range, but its soft modes become anisotropic and shift to incommensurate momenta. A promising direction for future study is to fully characterize these emergent critical modes. While a fermionic mean-field framework~\cite{1987AffleckHaldane, 1994Itoi,1997Itoi,2022PhysRevB.105.014435} could provide a qualitative understanding of these momentum shifts, computing the dynamical structure factors~\cite{2018BBQresponse,2020LowEnergyCritical,2022PhysRevB.105.014435} would offer direct, momentum-resolved access to the full excitation spectrum.

The intricate interplay among dimerized, nematic, and quadrupolar orders observed in 1D suggests a similarly rich landscape in the two-dimensional BBQK model on the honeycomb lattice. Although its global phase diagram has been studied~\cite{2DBBQK2023,2024PhysRevResearch,2024eightColorChiral}, the nature of the ground state in the positive biquadratic regime ($0 < \phi < \pi$) remains unsettled~\cite{2024PhysRevResearch}, and merits further investigation.

\begin{acknowledgments}
We thank Ian McCulloch for valuable discussions during a conference. This work was supported by the National Key R\&D Program of China
(No. 2022YFA1403000) and the National Natural Science Foundation of
China (Nos. 12274207 and 12174194). The numerical results in this work were obtained using the Julia package ITensor~\cite{ITensor}.
\end{acknowledgments}

\appendix

\section{\label{app:1}Haldane and LLRR phases}

  \begin{figure}[tb]
  \includegraphics[width=\linewidth]{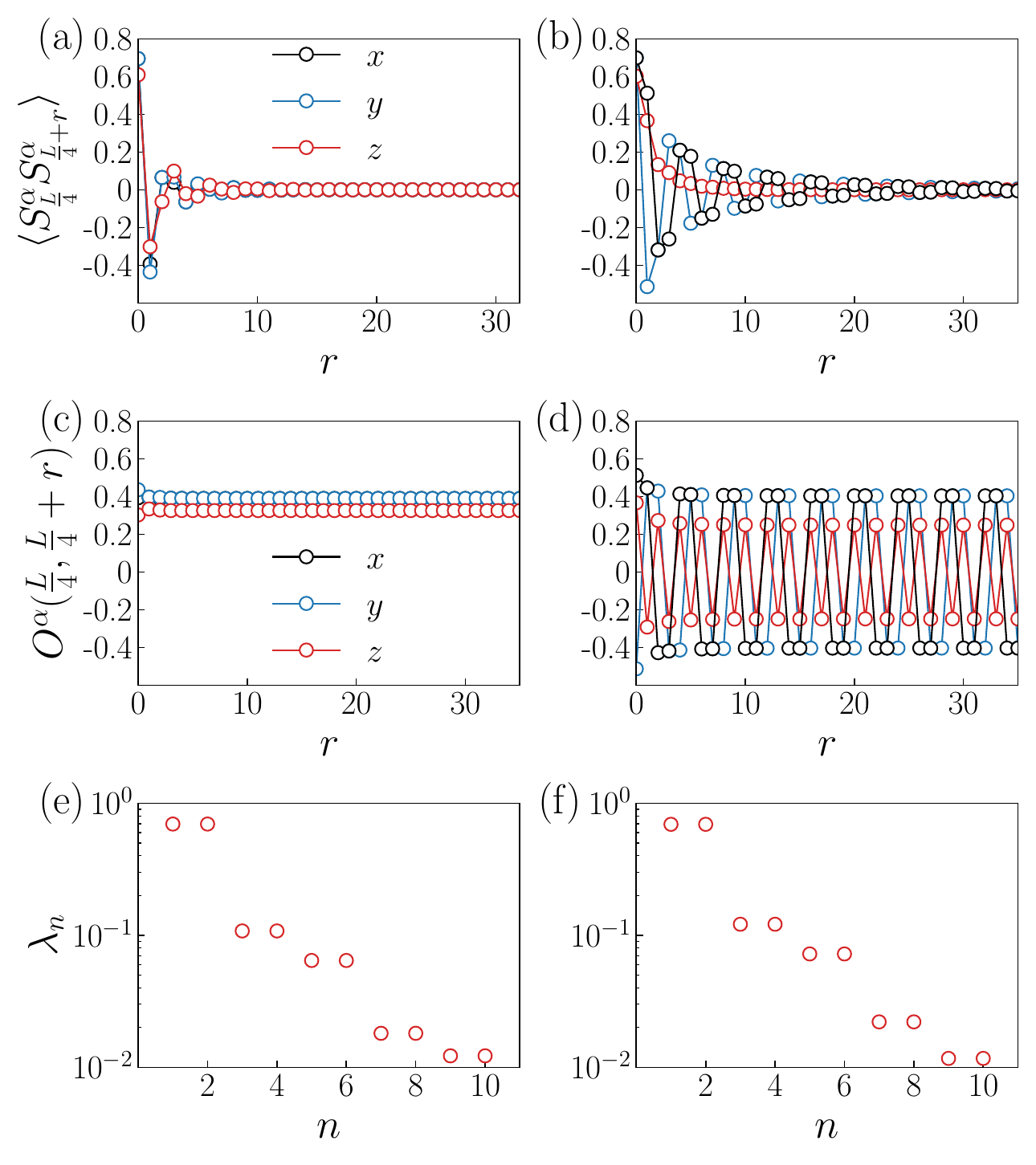}
  \caption{\label{fig:topo}
    (a,b) Spin correlation functions in (a) the Haldane phase at $(\theta,\phi) = (0.3\pi, 0.2\pi)$ and (b) the LLRR phase at $(\theta,\phi) = (0.38\pi, \pi)$.
    (c,d) String order parameters and (e,f) entanglement spectra computed respectively for the same parameters as (a) and (b).
    All data are from an open chain with length $L=129$ and spin-1/2 edge spins to suppress boundary effects.
  }
  \end{figure}

In this section, we briefly review the topological Haldane and LLRR phases, as discussed in earlier works~\cite{1983Haldane,1988Haldane,1992KTtransformation,Kennedy1992Z2Z2,Pollmann2010ESofSPT,WLYou2020HK,2023HKSIA}.

The Haldane phase is a prototypical example of an SPT phase in one dimension~\cite{1983Haldane,1988Haldane,1992KTtransformation,Kennedy1992Z2Z2,Pollmann2010ESofSPT}. Haldane conjectured that antiferromagnetic integer-spin chains are generically gapped, while half-integer chains remain gapless~\cite{1983Haldane,1988Haldane}. For the spin-1 BBQK chain with $(\theta,\phi) = (0.3\pi, 0.2\pi)$, the system lies in the Haldane phase, and this gapped nature is reflected in the exponentially decaying spin-spin correlations, as shown in Fig.~\ref{fig:topo}(a). The anisotropy between different spin components originates from the finite Kitaev interaction.
The nontrivial topology of the Haldane phase is reflected by symmetry-protected edge states, stabilized by time-reversal, bond-centered inversion, or dihedral spin rotation symmetry~\cite{1992KTtransformation,Kennedy1992Z2Z2,Pollmann2010ESofSPT}, and by a finite string order parameter [Eq.\eqref{eq:SO}], as shown in Fig.\ref{fig:topo}(c). This phase is adiabatically connected to the AKLT state~\cite{1988AKLT} [$(\theta, \phi) = (\pi/2, \arctan 1/3)$ in Fig.~\ref{Fig_PD}], which admits a valence-bond-solid description.
A hallmark of the Haldane phase is the double degeneracy of its entanglement spectrum, protected by the same symmetries~\cite{Pollmann2010ESofSPT}, as shown in Fig.~\ref{fig:topo}(e). Since the Kitaev interaction respects these symmetries, the Haldane phase remains stable as an SPT phase over a broad parameter regime.

The LLRR phase was first identified in the spin-1 Heisenberg-Kitaev chain~\cite{WLYou2020HK} with ferromagnetic bilinear interaction $J_1$, where it also exhibits nematic order~\cite{2023HKSIA}. Its name reflects the four-site correlation pattern observed in $\langle S^x_i  S^x_j\rangle$ and $\langle S^y_i  S^y_j\rangle$~\cite{WLYou2020HK}, as illustrated in Fig.~\ref{fig:topo}(b), where correlations exhibit a $(+,+,-,-)$ modulation over four consecutive bonds.
Despite these differences, the LLRR phase shares key SPT features with the Haldane phase. It displays an oscillatory yet finite string order [Fig.~\ref{fig:topo}(d)] and a doubly degenerate entanglement spectrum [Fig.~\ref{fig:topo}(f)], both protected by the hidden $\mathbb{Z}_2\times \mathbb{Z}_2$ symmetry, which remains intact in the presence of the Kitaev interaction.

\bibliography{bbqk}

\end{document}